\documentclass{iau}
\usepackage{graphicx}

\title[] 
{On the Information Content of\\Stellar Spectra}

\author[Jesper Schou]   
{Jesper Schou}

\affiliation{Max Planck Institute for Solar System Research\\
Max-Planck-Str. 2,
37191 Katlenburg-Lindau,
Germany\\
email: {\tt schou@mps.mpg.de}}

\pubyear{2013}
\volume{301}  
\pagerange{119--126}
\jname{Precision Asteroseismology: Celebration of the Scientific
Opus of
Wojtek Dziembowski}
\editors{W. Chaplin, J. Guzik, G. Handler \& A. Pigulski, eds}
\begin{document}

\maketitle

\begin{abstract}
With the increasing quality of asteroseismic observations
it is important to
minimize the random and systematic errors in mode parameter
estimates.
To this end it is important to understand how the oscillations
relate to the directly observed quantities, such as intensities and spectra,
and to derived quantities,
such as Doppler velocity.
Here I list some of the effects we need to take
into account and show an example of the impact of some of them.
\end{abstract}

\firstsection 
\section{Calculation of Mode Sensitivities}

To do precision asteroseismology, we need precise and accurate
observations, preferably for a large number of modes.
We thus need to understand how the oscillations
of a star relate to the observed quantities, such as spectra and
broad band intensity measurements.

Often the intensity perturbation caused by a mode is calculated by
multiplying a spherical harmonic with the limb darkening and integrating
over the disk. For a velocity observation,
a velocity projection factor is also included.
This neglects many physical, instrumental and data analysis effects and
can cause systematic errors, loss of S/N and
failure to make the best use of the observations.
It is thus planned to systematically attack these
problems, noting that significant work has been done in other
contexts (e.g. \cite[Zima 2006]{Z06}).
This is becoming increasingly important in view of the increased
quantity and quality of data soon to be available
from the SONG project (\cite[Grundahl, et al. 2009]{SONG09}).

For intensity observations some effects to consider include accurate limb darkening,
that the intensity perturbation is not necessarily given by the limb
darkening times the spherical harmonic (due to such effects as the
observing height
change with incidence angle), that there are non-adiabatic 
effects, that the mode phases depend on height
(\cite[Baldner \& Schou 2012]{BS12}), causing
center-to-limb phase shifts, the interaction of the oscillations
with convection causing spatially variable perturbations and so forth.
For spectrally-resolved observations (e.g. Doppler velocity) the
situation is further complicated by the
non-radial motion near the surface, change of height
of formation as a function of spectral line, position in the spectral line,
line broadening, convective blueshift etc.

\begin{figure}
\begin{center}
 \includegraphics[width=3.4in, angle=90]{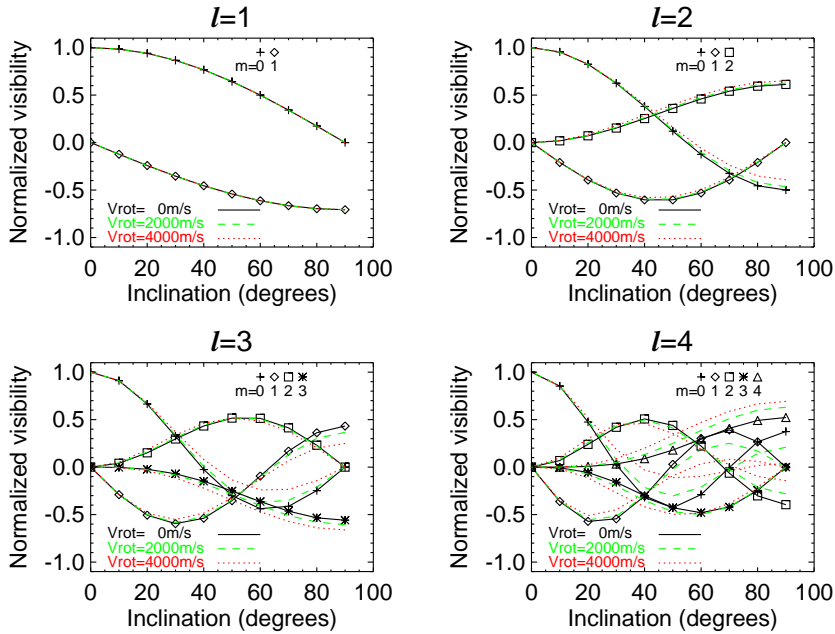} 
 \caption{Visibilities as a function of inclination angle, divided by the total visibility (summed in quadrature
over $m$). The results are labeled by the equatorial velocity.
}
   \label{fig1}
\end{center}
\end{figure}

To illustrate the importance of some of these effects, Fig. 1 shows
the relative mode visibilities for different values of the degree $l$,
and angular order $m$, as a function of inclination $i$.
To calculate these results, a snapshot of a solar MHD calculation
(courtesy of Bob Stein, using the Stagger code (\cite[Beeck, et al. 2012]{Beeck2012})
and the EOS and opacities from \cite{Nordlund82})
was used to synthesize the FeI 617.3nm line, as a function of
viewing angle. These profiles were then shifted according to
a solid body rotation law and averaged over the disk 
to create a reference line profile. The velocity pattern from
several low degree spherical harmonics were then added to the
velocity profile, integrated and cross correlated with the reference
profile to determine the mode sensitivity.

Fig. 1 shows that even a modest solar-like rotation causes a significant
change in the mode visibilities, especially at $l=3$ and $l=4$,
which could cause a significant mis-estimation of the inclination.
For unresolved modes it may also lead to mis-estimation of the
rotational splittings.
The zero velocity cases
follow the results of \cite{GS03}, as
expected. On the other hand many potentially
significant effects were not taken into account.
The oscillations were assumed to be radial at the surface,
constant as a function of height and unaffected by the granulation.
Also, no account was taken of differential rotation or
the possible variation of the phase or amplitude with height described
by \cite{BS12}.

\section{Conclusion}
It is clear that there are many effects we need to take into account
if we wish to accurately model stellar spectra and extract the maximum
information.
Fig. 1 only shows one example and the plan is
to systematically investigate all the relevant effects.

The author would like to thank Bob Stein, Bj\"{o}rn L\"{o}ptien, Aaron Birch,
Robert Cameron, Charles
Baldner, Regner Trampedach and Warrick Ball for insights and help with various aspects of the analysis.

\end{document}